\documentclass[conference]{IEEEtran}
\pdfoutput=1
\usepackage{amsmath}
\usepackage{algorithmic}
\usepackage{graphicx}
\usepackage{svg}
\usepackage[skip=3pt]{caption}
\usepackage{array}
\usepackage{cite}
\usepackage{hyperref}
\usepackage{xcolor,soul}
\usepackage{amssymb}
\usepackage{hhline}
\usepackage{siunitx}

\usepackage{booktabs, makecell, multirow, tabularx} 

\newcolumntype{C}{>{\centering\arraybackslash}X}
\newcolumntype{L}{>{\hsize=.4\hsize}C}
\newcolumntype{M}{>{\hsize=.35\hsize}C}
\newcolumntype{S}{>{\hsize=.25\hsize}C}

\usepackage{colortbl}
\definecolor{light-gray}{gray}{0.85}
\usepackage{caption} 
\usepackage{diagbox}
\usepackage{threeparttable}

\usepackage{pifont}
\usepackage{tikz}
\newcommand{\tikzxmark}{%
\tikz[scale=0.23] {
    \draw[line width=0.75,line cap=round] (0,0) to [bend left=6] (0.7,0.7);
    \draw[line width=0.75,line cap=round] (0.1,0.7) to [bend right=3] (0.65,0.05);
}}

\usepackage[ruled, linesnumbered, vlined]{algorithm2e}
\SetAlFnt{\small}
\SetAlgoNlRelativeSize{0}
\IncMargin{3pt}
\SetKwInOut{Input}{Input}
\SetKwInOut{Require}{Require}
\SetKwInOut{Output}{Output}

\newenvironment{para_noindent}{\setlength{\parindent}{0pt}}{}
\definecolor{moh_colour}{RGB}{255, 204, 204}

\definecolor{yuz_colour}{RGB}{191, 232, 255}

\definecolor{jord_color}{RGB}{66, 245, 99}

\begin{document}

\title{
    M4BRAM: Mixed-Precision Matrix-Matrix Multiplication in FPGA Block RAMs
}

        
\author{
    Yuzong Chen, Jordan Dotzel, Mohamed S. Abdelfattah \\
    \textit{Department of Electrical and Computer Engineering, Cornell University}  \\
    \{yc2367, dotzel, mohamed\}@cornell.edu
}

\maketitle

\begin{abstract}
Mixed-precision quantization is a popular approach for compressing deep neural networks (DNNs). 
However, it is challenging to scale the performance efficiently with mixed-precision DNNs given the current FPGA architecture and conventional accelerator dataflows. 
In this work, we enhance the FPGA's capability for accelerating mixed-precision DNNs by proposing M4BRAM, a novel compute-in-block RAM (BRAM) architecture that can compute mixed-precision matrix-matrix multiplication. 
On the precision side, M4BRAM supports a wide range of mixed-precision DNN configurations -- the weight precision can be 2/4/8 bits while the activation precision can vary from 2 to 8 bits. 
On the dataflow side, M4BRAM leverages a novel in-BRAM data duplication scheme to achieve high hardware utilization. Moreover, during M4BRAM computation, other FPGA resources can seamlessly access its data without the need for a separate buffer. 
Hence, unlike prior compute-in-BRAM proposals, M4BRAM can simultaneously perform mixed-precision computation and maintain full functionality as a memory unit to \textit{truly} complement the existing compute resources on FPGAs. 
Experiments show that adding M4BRAM to a tiled DNN accelerator can achieve an average speedup of 2.16$\times$ across various DNNs on the ImageNet classification task while incurring a negligible accuracy loss of $<$ 0.5\%. Compared to the same tiled accelerator that employs a prior compute-in-BRAM architecture, M4BRAM delivers 1.43$\times$ higher performance on average across various DNNs.

\end{abstract}


\section{Introduction}
Deep neural networks (DNNs) have demonstrated remarkable accomplishments in various important fields such as computer vision and natural language processing. 
Unfortunately, the scaling of compute performance and storage density following Moore's law has fallen far behind the growth of DNN model size \cite{cerebras}, which necessitates model compression to reduce the storage and computation cost of DNNs.
As one of the key compression techniques, quantization has been widely explored at both algorithmic \cite{BNN, TNN, postTrain_4bit, F8Net, mixedPrecisionDNN, autoQ, HAWQ, mpSurvey} and hardware levels \cite{stripes, bitFusion, hao, lowPrecIntelFPGA, elbNN, fabnet, filmQNN, msd, zPIM, lutCIM, programmableCIM, ccb, comefa, bramac}. 
Although uniform low-precision such as binary \cite{BNN}, ternary \cite{TNN}, and 4-bit \cite{postTrain_4bit} quantization can significantly reduce the model size and improve the computational throughput, it can lead to a noticeable reduction in model accuracy. 
To mitigate such accuracy loss, mixed-precision quantization \cite{mixedPrecisionDNN, autoQ, HAWQ, mpSurvey}, where both weights and activations can have different bit-widths, has emerged as a promising quantization approach. 


In the meanwhile, FPGAs have become an increasingly popular accelerator platform for DNNs due to their bit-programmability that allows customized precision and datapath \cite{fpgaConvnet}. 
Indeed, many mixed-precision DNN accelerators based on FPGAs have been proposed \cite{lowPrecIntelFPGA, elbNN, hao, filmQNN, msd}. These accelerators mainly rely on the digital signal processing (DSP) block to implement the multiply-accumulate (MAC) operation, the fundamental primitive in DNN computing. 
To further increase the peak MAC throughput of FPGAs for accelerating DNNs, recent works have suggested adding compute-in-memory (CIM) capability inside the FPGA block RAM (BRAM) \cite{ccb, comefa, bramac}. But these proposals either restrict DNN weights and activations to have the same precision \cite{bramac}, or have limited flexibility in feeding DSPs as a normal BRAM \cite{ccb, comefa, bramac}. For example, when the BRAM in \cite{ccb} and \cite{comefa} is configured to the CIM mode, it can no longer be accessed by DSPs. As a result, the available BRAM resources have to be partitioned into two groups -- one that performs CIM operations, and the other that feeds data to DSPs. This raises questions about the effective performance enhancement in realistic DNN accelerators on FPGAs. 


In this work, we address the above limitations to enhance the capability of FPGAs for accelerating mixed-precision DNNs by proposing a new CIM architecture called M4BRAM, which can compute \underline{M}ixed-precision \underline{M}atrix-\underline{M}atrix \underline{M}ultiplication in \underline{BRAM}. 
Compared to prior compute-in-BRAM proposals, M4BRAM has three novel contributions. First, it supports diverse mixed-precision DNNs whose weights can be 2/4/8 bits, and activations can vary from 2 to 8 bits. The MAC latency/throughput of M4BRAM can efficiently scale with lower activation/weight precision. 
Second, M4BRAM only occupies one BRAM port during in-memory computing, while another BRAM port is always available for the DSP to receive data. Therefore, M4BRAM can remain as a memory unit while performing computation to \textit{truly} complement the DSP. 
Third, M4BRAM employs a novel in-BRAM data duplication scheme that exploits DNN computational parallelism through both weight-sharing and activation-sharing to achieve high hardware utilization. 

To demonstrate the flexibility and efficiency of M4BRAM in accelerating mixed-precision DNNs, we propose a heterogeneous tiled accelerator with M4BRAM functioning as a bit-serial engine and the DSP operating as a bit-parallel engine. The heterogeneous tiled accelerator significantly outperforms the baseline tiled accelerator without M4BRAM. 
Through activation quantization, it can deliver an average speedup of 2.16$\times$ across various DNNs while incurring a negligible accuracy loss of $<$ 0.5\% compared to the floating-point model on ImageNet classification.
The performance gains further increase by adopting intra-layer weight quantization. 
Finally, compared to the heterogeneous accelerator that employs an existing compute-in-BRAM architecture, M4BRAM can provide 1.43$\times$ higher performance on average across various DNNs. 


\section{Related Works} \label{relatedWorks}

    \subsection{FPGA Acceleration of Mixed-Precision DNNs} \label{fpga_for_mp_dnn}
    The FPGA's bit-level programmability makes it a competitive acceleration platform for mixed-precision DNNs. Colangelo \textit{et al.} \cite{lowPrecIntelFPGA} assigned two separate precisions to weights and activations, respectively, and quantified the effects on accuracy, throughput, and FPGA resource utilization.
    Wang \textit{et al.} \cite{elbNN} and Sun \textit{et al.} \cite{filmQNN} investigated accelerating inter-layer and intra-layer mixed-precision DNNs on FPGA, respectively. 
    Another recent study, MSD \cite{msd}, proposed quantizing DNN weights using two number representations, which can take advantage of the heterogeneous computing resources on FPGA to perform bit-serial and bit-parallel computation simultaneously. 

    To efficiently utilize the high-precision DSP multiplier for mixed-precision DNNs, these FPGA accelerators commonly employ DSP-packing \cite{dspPacking4} to combine multiple low-precision multiplications into one DSP. Unlike these works based on the existing FPGA architecture, our proposed M4BRAM is a new BRAM architecture that can complement DSP-packing to further improve the performance of mixed-precision DNNs. 

    \subsection{Compute-In-Memory for Mixed-Precision DNNs}
    CIM has emerged as a new computing paradigm that achieves higher performance and energy efficiency than conventional Von-Neumann architectures by performing computation closer to the data. Although many CIM variants for mixed-precision DNNs have been proposed as ASICs \cite{zPIM, lutCIM, programmableCIM}, they suffer from limited precision support, such as \cite{programmableCIM} that only supports 1-/2-bit DNN inference and \cite{zPIM} that allows arbitrary weight precision but fixed 16-bit activation precision. 
    
    On the other hand, two recent BRAM-based CIM architectures, CCB \cite{ccb} and CoMeFa \cite{comefa}, can compute with any precision using bit-serial arithmetic. However, during the CIM mode, their BRAM cannot support double-buffering 
    which is a widely used technique in FPGA accelerators for DNNs \cite{hao, lowPrecIntelFPGA, elbNN, filmQNN, msd}. 
    A later compute-in-BRAM work, BRAMAC \cite{bramac}, performs CIM operations in a small dummy BRAM array that is decoupled from the main BRAM array. It proposes two variants, BRAMAC-2SA with two synchronous dummy BRAM arrays and BRAMAC-1DA with one double-pumped dummy BRAM array. In addition, the dummy BRAM array is controlled by an embedded finite-state machine (eFSM) to enable double-buffering. Nevertheless, BRAMAC requires weights and activations to have the same precision in 2-/4-/8-bit, limiting its applicability to only uniform-precision DNNs. 
    Our proposed M4BRAM extends some concepts from BRAMAC to support variable activation precision from 2 to 8 bits with linearly scaled MAC latency. 
    Moreover, M4BRAM simplifies interoperability with DSPs and improves hardware utilization through a novel data duplication scheme that can extract DNN computation parallelism from both weight-sharing and activation-sharing.

\section{Motivation} \label{motivation}



    \begin{figure}
        \centering
        \includegraphics[width=1\linewidth]{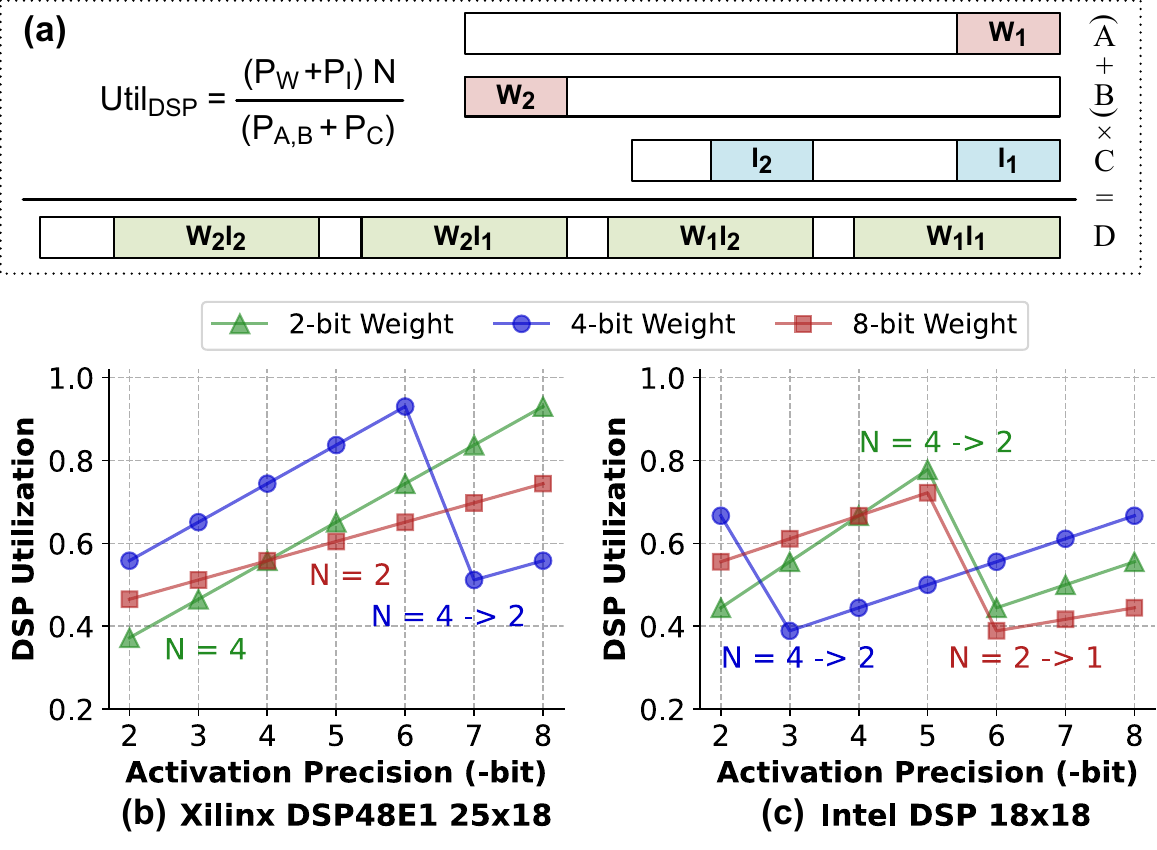}
        \caption{(a) Packing 4 multiplications onto one DSP. The achievable packing factor (N) and resulting DSP utilization ($Util_{DSP}$) depend on weight precision ($P_{W}$), activation precision ($P_{I}$), and the DSP multiplier size ($P_{A,B}$ and $P_{C}$), which can vary across different FPGA vendors: (b) Xilinx 25$\times$18-bit and (c) Intel 18$\times$18-bit.}
        \label{fig1_dsp_packing}
    \end{figure}
    
    \subsection{DSP-Packing is Sub-Optimal for Mixed-Precision DNNs} \label{DSP_Packing_Suboptimal}
    
    As described in Section \ref{fpga_for_mp_dnn}, DSP-packing is widely adopted in FPGA accelerator design. 
    Fig. \ref{fig1_dsp_packing}(a) shows an example of packing four multiplications between two weights ($W_{1}$ and $W_{2}$) and two activations ($I_{1}$ and $I_{2}$) onto one DSP following the method in \cite{dspPacking4}. In general, the achievable packing factor and resulting DSP utilization depend on the operand precision and the DSP multiplier size, with the latter varying across different FPGA vendors. 
    
    Fig. \ref{fig1_dsp_packing}(b) and (c) show the DSP utilization of packing mixed-precision multiplications on Xilinx and Intel DSPs, respectively. 
    The decreasing part in a line reflects a 2$\times$ reduction in the achievable packing factor. First, given the same weight precision, reducing the activation precision cannot increase MAC throughput in most cases since the packing factor does not change, although one purpose of quantization is to scale the performance proportionally with reduced activation precision \cite{stripes}. Second, given the same activation precision, reducing the weight precision by 2$\times$ also does not change the packing factor in many cases, although theoretically, a 2$\times$ throughput improvement is expected and can be achieved in some ASIC accelerators \cite{bitFusion}. Based on these observations, it is necessary to design new FPGA blocks that can scale the performance with fine-grained quantization to get \textit{measurable} hardware speedup for diverse mixed-precision DNNs. 

    
    \subsection{CIM Should Complement DSPs} \label{CIM_Complement_DSP}

    \begin{figure}
        \centering
        \includegraphics[width=1\linewidth]{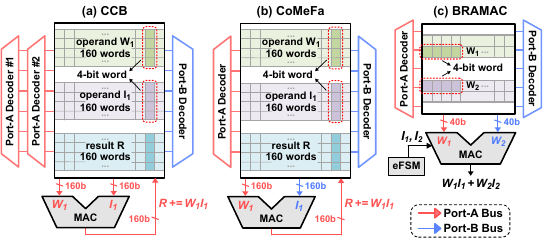}
        \caption{Top-level dataflow of (a) CCB, (b) CoMeFa, (c) BRAMAC. The data word is assumed to be 4-bit.  }
        \label{fig2_top_level_dataflow}
    \end{figure}

    To fully exploit the potential of CIM on FPGA, BRAM should remain as a memory unit to feed the DSPs while performing CIM operations. Unfortunately, a pure bit-serial arithmetic approach (similar to CCB\cite{ccb} and CoMeFa\cite{comefa}) requires a transposed data layout, where each data word is stored along the BRAM column as shown in Fig. \ref{fig2_top_level_dataflow} (a) and (b), respectively. Hence, when a BRAM block is operating in the CIM mode, a DSP can no longer access its data, therefore requiring a separate BRAM block configured as normal memory to feed the DSP.
    This can be wasteful and inefficient, especially when the data within the BRAM can be reused by more than one type of processing element as in heterogeneous DNN accelerators \cite{msd}.
    
    BRAMAC~\cite{bramac} solves the data layout issue by storing data words along the BRAM row, as shown in Fig. \ref{fig2_top_level_dataflow}(c). However, when running in CIM mode, its two BRAM ports are both occupied by the in-BRAM MAC unit. 
    Therefore, a DSP has to either receive data from a separate BRAM configured as normal memory or read the same operand being used by the in-BRAM MAC unit,
    resulting in constrained dataflow and limited parallelism. To overcome these challenges, a new CIM architecture that allows random access by the DSP while performing in-BRAM computation is desirable. 

\section{M4BRAM Architecture and Dataflow} \label{M4BRAM_Architecture}

    \subsection{Overview} \label{M4BRAM_Overview}

\begin{figure}
        \centering
        \includegraphics[width=1\linewidth]{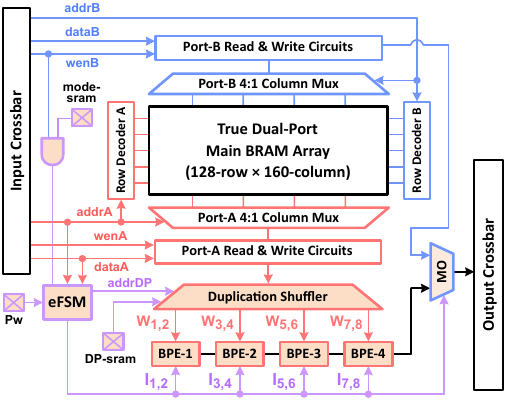}
        \caption{Top-level architecture of M4BRAM modified from the Intel M20K BRAM. The added circuit blocks are highlighted in orange. 
        }
        \label{fig4_m4bram_architecture}
    \end{figure}

    
    Fig. \ref{fig4_m4bram_architecture} shows the top-level architecture of M4BRAM based on the Intel M20K BRAM \cite{bramIntel} with added circuit blocks colored in orange. The routing interface (i.e., input and output crossbar) of M4BRAM is the same as that of M20K. The configuration SRAM cell, \textbf{mode-sram}, allows M4BRAM to operate in the memory mode or the compute mode. During the memory mode, M4BRAM is identical to a conventional BRAM with configurable depth and data width. During the compute mode, M4BRAM can perform two MAC operations simultaneously, which is called a MAC2 operation \cite{bramac} and defined as $P = (W_{1}I_{1} + W_{2}I_{2})$. The MAC2 operation is performed by 4 in-BRAM processing elements (BPEs) that are decoupled from the main BRAM array. This approach of decoupling computation from the main BRAM array is similar to that of BRAMAC. However, unlike BRAMAC, M4BRAM only occupies one port (port-A) of the main BRAM array to communicate with BPEs. A new duplication shuffler can reorder and replicate the weight vector read from the main BRAM array, which allows the 4 BPEs to compute 4 independent MAC2 operations, respectively. The rest of this section details M4BRAM's novel dataflow and new circuit blocks that make it much easier to be integrated within a DNN accelerator, surpassing prior works in terms of flexibility and hardware utilization.

    \subsection{In-BRAM Compute Dataflow}
    
    In the compute mode, M4BRAM is automatically configured as a simple dual-port memory with a depth of 512 and a data width of 32 bits, which is the BRAM data width supported by both Intel~\cite{bramIntel} and Xilinx\cite{bramXilinx}.
    The main BRAM array's port-A and port-B are configured as write and read ports, respectively. 
    Conventionally, the write-enable signal of the read port now becomes ``don't care". This idle signal, however, is reused in M4BRAM to dynamically indicate a CIM instruction. If the write-enable signal \textbf{wenB} of the read port-B is asserted, then the write port-A's data \textbf{dataA} and address \textbf{addrA} will be treated as a CIM instruction and sent to the eFSM. This controls the 4 BPEs, each can compute an independent MAC2 operation between weights \textbf{W} and activations \textbf{I}. The weight precision \textbf{Pw} is pre-defined in configuration SRAM while the activation precision is run-time configurable through the CIM instruction as described later in Section \ref{M4BRAM_Instruction}.

    To start a MAC2 operation, a 32-bit weight vector consisting of multiple low-precision weight elements is copied from the main BRAM array to 4 BPEs through port-A's read circuits. The duplication shuffler can select a subset of the weight vector and duplicate it multiple times depending on the duplication factor stored in configuration SRAM \textbf{DP-sram}. The activation vector is sent from outside (e.g., an activation buffer) to the eFSM through \textbf{dataA}, and then fed to the 4 BPEs. 
    Since the read port-B is not occupied by BPE and eFSM during the entire MAC2 operation, it can continue to send the main BRAM array's data to other logic resources such as DSP. After M4BRAM finishes computing, port-B is used to read out the final result from BPE. The 2-to-1 mux \textbf{MO} selects the output data between BPE and the main BRAM array. 

    \subsection{In-BRAM Processing Element (BPE)} \label{M4BRAM_BPE}
    
    The circuit design of BPE is based on the dummy BRAM array of BRAMAC \cite{bramac} but with a different physical geometry. In BRAMAC, the dummy BRAM array has a physical geometry of 7 rows $\times$ 160 columns. It can multiply the same activation with five 8-bit, ten 4-bit, or twenty 2-bit weights. 
    In M4BRAM, however, each BPE contains a dummy array with 7 rows $\times$ 32 columns\footnote{As described later in Section \ref{M4BRAM_Variants}, we also propose a second M4BRAM variant with a 7-row $\times$ 64-column dummy BRAM array in each BPE.} that can multiply the same activation with one 8-bit, two 4-bit, or four 2-bit weights. 
    
\begin{figure}
        \centering
        \includegraphics[width=1\linewidth]{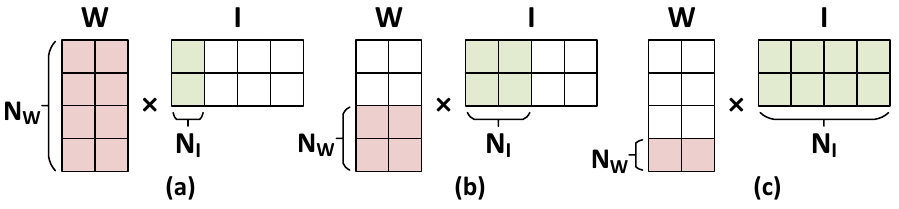}
        \caption{Different parallelism configurations supported by M4BRAM for 8-bit weight precision: (a) $N_{W}$=4, $N_{I}$=1; (b) $N_{W}$=2, $N_{I}$=2; (c) $N_{W}$=1, $N_{I}$=4. }

        \label{fig5_m4bram_parallelism}
    \end{figure}

    Another key difference between M4BRAM BPE and BRAMAC is the supported parallelism configuration defined by $N_{W}$ and $N_{I}$, where $N_{W}$ represents the number of weights multiplied by one activation and $N_{I}$ represents the number of activations multiplied by one weight. 
    BRAMAC only exploits the computation parallelism of DNNs through activation-sharing, i.e., it can only multiply the same activation with many weights by keeping $N_{I}$ = 1 and scaling $N_{W}$. On the other hand, the 4 BPEs in M4BRAM can receive different activations and multiply them with the same weight to also exploit weight-sharing. Fig. \ref{fig5_m4bram_parallelism} shows three different parallelism configurations supported by M4BRAM when the weight precision is 8 bits. The color-shaded area in each configuration represents 4 MAC2 that can be computed by 4 BPEs, respectively. 


    The reasoning behind supporting different parallelism configurations is that scaling only $N_{W}$ can lead to low hardware utilization for DNNs that do not exhibit enough activation-sharing parallelism, as demonstrated in a recent study from Intel \cite{dlaNew}. The study compares the hardware utilization of a tiled DNN accelerator across various parallelism configurations and DNNs, and shows that finding a balanced parallelism configuration based on the DNN's topology can significantly improve the hardware utilization and performance. Hence, compared to BRAMAC which only supports one parallelism configuration (i.e., $N_{I}$ = 1), M4BRAM has the potential to achieve higher performance by flexibly adjusting its parallelism configuration for specific DNNs. Section \ref{M4BRAM_vs_BRAMAC} will quantify the benefits of M4BRAM over BRAMAC for accelerating various DNNs.
     


    \subsection{Duplication Shuffler} \label{M4BRAM_DS}

    \begin{figure}
        \centering
        \includegraphics[width=1\linewidth]{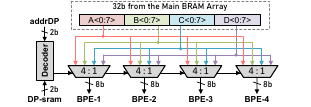}
        \caption{Duplication shuffler. 
        }
        \label{fig6_duplication_shuffler}
    \end{figure}
    
    In conventional FPGA accelerators with the compute engine consisting of only DSP and soft logic, the same weight read from BRAM can be routed to different processing elements and multiplied by different activations to effectively increase $N_{I}$. In other words, the same weight must be \textit{duplicated} in different processing elements to enable weight-sharing. 
    To realize such weight duplication in M4BRAM, we propose a duplication shuffler as shown in Fig. \ref{fig6_duplication_shuffler}. The duplication shuffler  is composed of four 4-to-1 mux, all controlled by a decoder. The 32-bit weight vector read from the main BRAM array is divided into 4 slices (A, B, C, D) and sent to all 4-to-1 muxes.  
    The decoder receives a 2-bit address \textbf{addrDP} to select a weight slice and duplicates it 1, 2, or 4 times depending on the duplication factor \textbf{DP-sram} (i.e., $N_{I}$). For example, if \textbf{DP-sram} = 1, then \textbf{addrDP} is ignored and the 4 BPEs will receive A, B, C, D, respectively, which gives the parallelism configuration shown in Fig. \ref{fig5_m4bram_parallelism}(a). On the other hand, if \textbf{DP-sram} = 4, then an 8-bit weight slice will be selected based on \textbf{addrDP} and broadcast to all BPEs, which gives the parallelism configuration shown in Fig. \ref{fig5_m4bram_parallelism}(c).

    \subsection{CIM Instruction Design} \label{M4BRAM_Instruction}
    
    \begin{figure}
        \centering
        \includegraphics[width=1\linewidth]{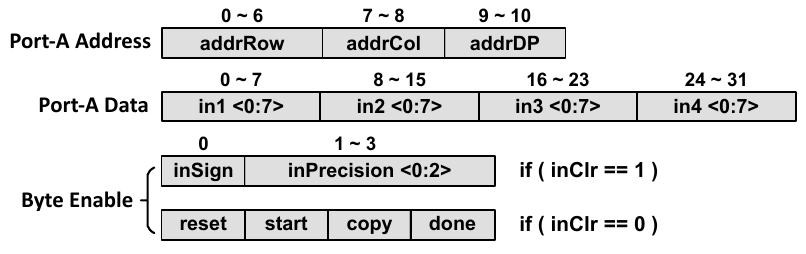}
        \caption{M4BRAM's CIM instruction format.}
        \label{fig7_m4bram_instruction}
    \end{figure}

    \begin{figure*}
        \centering
        \includegraphics[width=1\linewidth]{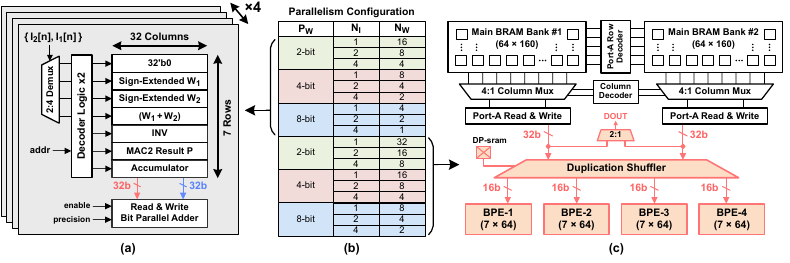}
        \caption{M4BRAM's (a) BPE diagram; (b) Supported parallelism configurations ($N_{I}$ and $N_{W}$) under different weight precision ($P_{W}$); \\ 
        (c) Architecture of M4BRAM-L. The DOUT control can come from the most significant bit of the read address.}
        \label{fig8_m4bram_variant}
    \end{figure*}

     Fig. \ref{fig7_m4bram_instruction} shows M4BRAM's CIM instruction format. Before starting a MAC2 operation, the eFSM needs 2 cycles to receive 2 CIM instructions, which contain information for 2 weight/activation vectors, respectively. The port-A address contains row and column addresses (\textbf{addrRow} and \textbf{addrCol}) of a weight vector that needs to be copied from the main BRAM array to BPE. It also includes a 2-bit \textbf{addrDP} to control the duplication shuffler. The port-A data contains 4 input activations to be multiplied by the weight vector in 4 BPEs. Additional control information is encoded in each BRAM's 4-bit byte-enable signal which is supported by both Intel and Xilinx BRAMs. The \textbf{inClr} can be any M20K control signal that is unused by the main BRAM array during the compute mode. When \textbf{inClr} is activated, the input activation sign and precision can be changed to support layer-wise mixed activation precision. When \textbf{inClr} is deactivated, the byte-enable will contain flags to control the MAC2 operation.

    \subsection{Mixed-Precision Support} \label{MixedPrecision_Support}
    The proposed CIM instruction design allows M4BRAM's BPE to compute MAC2 with 2-/4-/8-bit weights and 2- to 8-bit activations chosen dynamically. Fig. \ref{fig8_m4bram_variant}(a) shows the diagram of a BPE consisting of a 7-row $\times$ 32-column dummy BRAM array and peripheral circuits. The BPE performs computation on sign-extended weight vectors 
    and uses a look-up table approach to generate the partial sum \cite{lutCIM}. In every cycle, based on the received two activation bits $\{I_{2}[n], I_{1}[n]\}$, the partial sum is chosen from one of the first four BRAM rows and added to the MAC2 result $P$. The $INV$ row is used to store a temporary inverted partial sum if the activation is a signed number. The final MAC2 result $P$ is added to the last BRAM row for accumulation. More details on the above bit-serial MAC2 dataflow, as well as the circuit design of different components, are detailed in \cite{bramac}. 
    
    With bit-serial processing of activations, the MAC2 operation in M4BRAM takes $(n + 2)$ cycles for $n$-bit activation when the BPE is synchronous with the main BRAM array, and this latency can be further reduced to $(n/2 + 2)$ by double-pumping the BPE with a 2$\times$ main BRAM clock frequency \cite{bramac}. Furthermore, a lower weight precision improves MAC2 throughput because the 32-bit weight vector copied from the main BRAM array can contain more elements.
    As a result, M4BRAM can provide \textit{measurable} hardware speedup for diverse mixed-precision DNNs through both activation quantization and weight quantization. 


    \subsection{M4BRAM Variants} \label{M4BRAM_Variants}
    To explore the trade-off between the MAC throughput gain and the area overhead, we propose two M4BRAM variants, called M4BRAM-S and M4BRAM-L, whose BPEs contain small and large dummy BRAM arrays, respectively. The BPE of M4BRAM-S has the same structure as the one described in Section \ref{MixedPrecision_Support}. Fig. \ref{fig8_m4bram_variant}(b) shows the parallelism configurations supported by M4BRAM-S for different weight precision. Reducing the weight precision allows multiplying the same activation with more weights, leading to higher $N_{W}$.
    This variant has a smaller area overhead and offers moderate MAC throughput gain. 
    
    Fig. \ref{fig8_m4bram_variant}(c) shows the architecture of M4BRAM-L whose BPE contains a large dummy BRAM array with 7 rows and 64 columns. Compared to M4BRAM-S, M4BRAM-L can achieve 2$\times$ higher weight-sharing parallelism as shown in Fig. \ref{fig8_m4bram_variant}(b). But it also requires 2$\times$ read bandwidth to copy more weights from the main BRAM array. To achieve this, we apply memory banking to divide the main BRAM array into two 64-row $\times$ 160-column banks so that each can provide a 32-bit weight vector. Note that the read bandwidth is increased only internal to the BRAM, an additional 2-to-1 mux selects between two banks during normal BRAM access. 
    

    \subsection{Heterogeneous Accelerator with M4BRAM and DSP} \label{Heterogeneous_Architecture}
    
    During a MAC2 operation, the BPE and eFSM of M4BRAM only occupy the main BRAM array's write port (for 2 cycles) to receive operands, while the DSP can continue to randomly access the main BRAM array through the read port. Hence, the BPE and DSP can receive data from a shared M4BRAM cache and complement each other in a heterogeneous fashion, where the BPE is a bit-serial engine with its latency proportional to the activation precision, and the DSP is a bit-parallel engine with fixed 1-cycle latency. 
    
\begin{figure}
        \centering
        \includegraphics[width=1\linewidth]{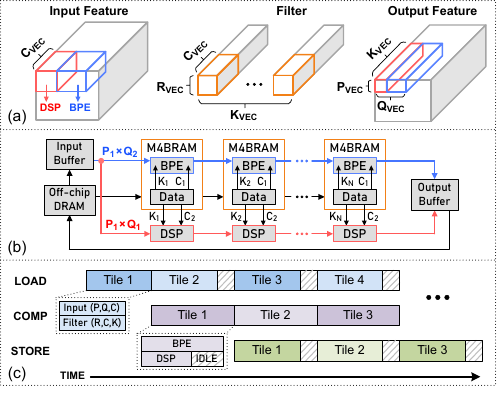}
        \caption{(a) Partitioning a CNN into tiles along different dimensions in DLA. The tile is distributed to BPE and DSP along $Q_{VEC}$. (b) Proposed heterogeneous DLA. 
        (c) Inter-tile pipelined dataflow.
        }
        \label{fig9_heterogeneous_architecture}
    \end{figure}

    To illustrate how the BPE and DSP can coordinate in a tiled accelerator, we use Intel's Deep Learning Accelerator (DLA) \cite{dlaNew} as an example. Fig. \ref{fig9_heterogeneous_architecture}(a) shows how a CNN can be partitioned into tiles along different dimensions in DLA, where $C_{VEC}/K_{VEC}$ represents the tile size along the input/output channel, $R_{VEC}$ represents the tile size along the filter height and width (assuming square filter), and $P_{VEC}/Q_{VEC}$ represents the tile size along output height/width. 
    In order for the BPE and DSP to work in parallel, the workload can be distributed along $Q_{VEC}$ so that they will receive/compute different inputs/output features (indicated by blue and red cubes in Fig. \ref{fig9_heterogeneous_architecture}(a)).
    It is also possible to partition the workload along other dimensions, which is left for future work. 

    Fig. \ref{fig9_heterogeneous_architecture}(b) shows the proposed heterogeneous accelerator (Hetero-DLA) that enhances DLA by using M4BRAM to store data. The input buffer sends two sets of input features to the BPE and DSP, respectively. The filter data stored in M4BRAM's main BRAM array can be randomly accessed by both the BPE and DSP. For example, if the DSP computes faster, it can start to compute the next tile, \textbf{C2}, along the input channel while the BPE is still computing on \textbf{C1}. The final results from the two engines are sent to the output buffer. 
    
    Existing tiled CNN accelerators commonly apply double-buffering to reduce or hide off-chip DRAM access latency \cite{lowPrecIntelFPGA, elbNN, hao, filmQNN, msd}. Double-buffering allows an accelerator to pipeline three stages: loading inputs and filters, computing convolution, and storing outputs, as illustrated in Fig. \ref{fig9_heterogeneous_architecture}(c). 
    Similar to BRAMAC, M4BRAM supports double-buffering during the compute mode -- when the eFSM is not receiving a CIM instruction, the main BRAM array's write port is free and can be used to load the next tile. 
    During convolution, the tile latency is determined by the slower latency between the BPE and DSP. In addition, when the BPE completes a dot product, the result needs to be read out through the main BRAM's read port, forcing the DSP to stall for 4 cycles and 8 cycles in M4BRAM-S and M4BRAM-L, respectively given a 32-bit BRAM data width. 
    Fortunately, this stall overhead can be amortized over many MAC2 operations in the BPE since the dot product size is usually large in real-world CNNs. 
    For example, our performance simulator (discussed in Section \ref{Experimental_Setup}) shows that such DSP stalls contribute to an average of only 4.8\% of the total execution time for VGG-16 with 8-bit weights and the activation varying from 4 to 8 bits. 

\section{Evaluation} \label{Evaluation}

    \subsection{Experimental Setup} \label{Experimental_Setup}

    \begin{para_noindent}
    \textbf{M4BRAM modeling:} We use COFFE \cite{COFFE} with 22-nm predictive technology to model the area and delay of the baseline M20K as well as new M4BRAM circuits. For M4BRAM's eFSM, we write SystemVerilog and synthesize it using Synopsys Design Compiler with TSMC 28-nm technology and scale the reported area to 22-nm technology. 

    \textbf{DNN benchmarks:} We choose AlexNet, VGG-16, ResNet-18, ResNet-34, and one self-attention module from the base Vision Transformer (ViT-Base) \cite{vit}. For DNNs used in the mixed-precision training experiments, we train them on ImageNet classification. The baseline 32-bit floating-point (FP32) models are taken from PyTorchCV \cite{pytorchcv} and quantized to fixed-point using uniform symmetric quantization. The quantization clipping thresholds are determined by minimizing the mean absolute error on the original weights and activations, where activation statistics are estimated with a large random training batch. For experiments involving intra-layer weight quantization, the weights are partitioned into two slices along the output dimension and then quantized individually with the above method. The models are then fine-tuned using the default Adam optimizer with a learning rate of 1e-5 for 20 epochs following a cosine decay learning rate schedule.
    
    \textbf{Baseline and enhanced accelerator:} We choose two baseline FPGAs from Intel's Stratix-10 family: GX400 and GX650 (Table \ref{baseline_fpga}). We apply Hetero-DLA described in Section \ref{Heterogeneous_Architecture} to evaluate the DNN benchmarks. We compare the performance of Hetero-DLA to that of DLA, which has normal BRAM instead of M4BRAM. The matrix multiplication operation in the self-attention module of ViT-Base is converted to 1D convolution \cite{dlaNew}. 
    To maximize each accelerator's performance, we develop a design space exploration tool to find the optimal tiling configuration for every DNN, following a similar approach used in prior FPGA accelerator works \cite{dlaNew, filmQNN}. The optimization target is set to $perf \times (perf/area)$ to balance the performance and area cost.
    We build a cycle-accurate simulator to obtain the latency of the DSP and BPE based on a given tiling configuration that splits the workload distribution between the M4BRAM and DSP along the $Q_{VEC}$ dimension.

    \end{para_noindent}


    \begin{table}
      \centering
      \caption{Properties of the baseline Stratix-10 FPGAs}
      \renewcommand{\arraystretch}{1.2}
      
      \begin{threeparttable}
        \begin{tabular}{ | m{2cm}<{\centering} |m{1.2cm}<{\centering} S[table-column-width=1.2cm] | m{1.2cm}<{\centering} S[table-column-width=1.2cm] |}
            \hhline{~|*4-}  
                \multicolumn{1}{c|}{} &
                \multicolumn{2}{c|}{\cellcolor{light-gray} GX400} &
                \multicolumn{2}{c|}{\cellcolor{light-gray} GX650} \\
            \hline
                {Resources} & {Count} & {Area(\%)\tnote{1}}  & {Count} & {Area(\%)\tnote{1}}  \\
            \hline
                Logic Block   & 12816 & 55.6  & 20736 & 54.7  \\
                DSP           & 648   & 15.7  & 1152  & 17.0  \\
                M20K BRAM     & 1537  & 28.7  & 2489  & 28.3  \\
            \hline
        \end{tabular}
        
        \begin{tablenotes}
          \item[1] Calculated based on the normalized area metrics in \cite{tensorSlice}. 
        \end{tablenotes}
      \end{threeparttable}

      \vspace{-5pt}
      \label{baseline_fpga}
    \end{table}

    \begin{table*}
      \centering
      \caption{Comparison between M4BRAM and prior compute-in-BRAM architectures}
      \renewcommand{\arraystretch}{1.2}
      
        \begin{tabular}{ | m{4.2cm}<{\centering} | m{1.2cm}<{\centering} | m{1.2cm}<{\centering} | m{1.2cm}<{\centering} | m{1.2cm}<{\centering} | m{1.2cm}<{\centering} | m{1.2cm}<{\centering} | m{1.2cm}<{\centering} |}
            \hhline{*8{-}}
                \multicolumn{1}{|c|}{\cellcolor{light-gray} Property} &
                \multicolumn{1}{c|}{\cellcolor{light-gray} CCB} &
                \multicolumn{2}{c|}{\cellcolor{light-gray} CoMeFa} &
                \multicolumn{2}{c|}{\cellcolor{light-gray} BRAMAC} &
                \multicolumn{2}{c|}{\cellcolor{light-gray} M4BRAM} \\
            \hline
                {Variant} &  -- & {D} & {A} & {1DA} & {2SA} & {S} & {L}  \\
            \hline
                {\# of dummy arrays} & {--} & {--} & {--} & {1} & {2} & {4} & {4}  \\
            \hline
                {Dummy array size} & {--} & {--} & {--} & {7$\times$160} & {7$\times$160}  & {7$\times$32} & {7$\times$64}  \\
            \hline
                {Weight precision (-bit)} & {Arbitrary} & {Arbitrary} & {Arbitrary} & {2, 4, 8} & {2, 4, 8}  & {2, 4, 8} & {2, 4, 8}  \\
            \hline
                {Activation precision (-bit)} & {Arbitrary} & {Arbitrary} & {Arbitrary} & {2, 4, 8} & {2, 4, 8}  & {2 -- 8} & {2 -- 8}  \\
            \hline
                {Weight-sharing factor $N_{I}$} & {1} & {1} & {1} & {1} & {2}  & {1, 2, 4} & {1, 2, 4}  \\
            \hline
                {Use transposed data layout}  & {\checkmark} & {\checkmark} & {\checkmark} & {\tikzxmark} & {\tikzxmark} & {\tikzxmark} & {\tikzxmark}  \\
            \hline
                {Allow DSP access during CIM}  & {\tikzxmark} & {\tikzxmark} & {\tikzxmark} & {\tikzxmark} & {\tikzxmark} & {\checkmark} & {\checkmark}  \\
            \hline
                {Support mixed-precision}  & {\checkmark} & {\checkmark} & {\checkmark} & {\tikzxmark} & {\tikzxmark} & {\checkmark} & {\checkmark}  \\
            \hline
                {Support multi-pumping} & {\tikzxmark} & {\tikzxmark} & {\checkmark} & {\checkmark} & {\tikzxmark}  & {\checkmark} & {\checkmark}  \\
            \hline
                {\# of occupied main BRAM ports} & {Two} & {Two} & {Two} & {Two} & {Two} & {One} & {One}  \\
            \hline
                {M20K area overhead} & {16.8\%} & {25.4\%} & {8.1\%} & {16.9\%} & {33.8\%}  & {19.6\%} & {33.4\%}  \\
            \hline
        \end{tabular}
      \vspace{5pt}
      \label{m4bram_vs_prior}
    \end{table*}

    \subsection{M4BRAM Area and Frequency} \label{M4BRAM_Area_Frequency}


    \begin{figure*}
        \centering
        \includegraphics[width=1\linewidth]{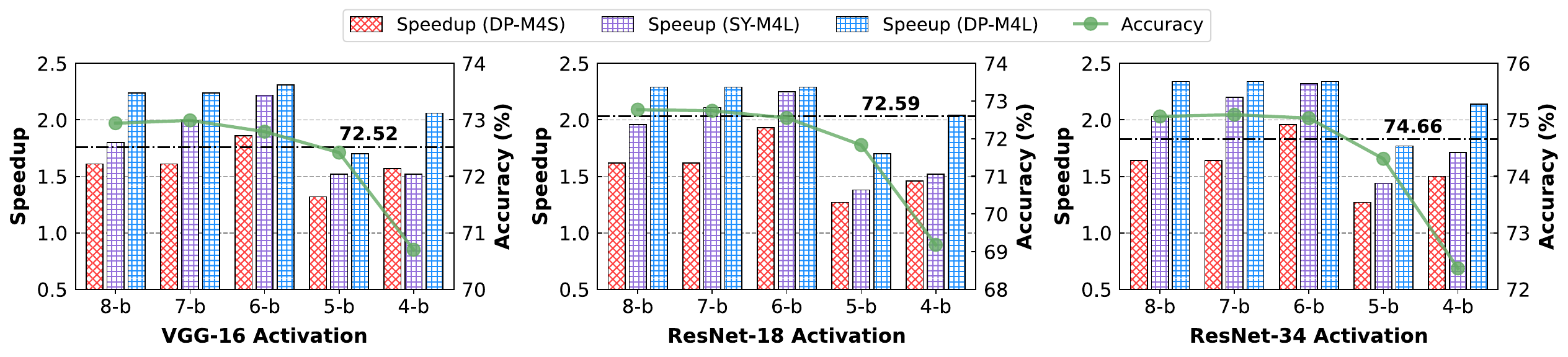}
        \caption{Model accuracy and speedup over DLA for various DNNs. The weight precision is 8-bit. The dash-dotted black line represents a 0.5\% Top-1 accuracy loss compared to the baseline FP32 model. }
        \label{fig11_m4bram_mixed_activation_performance}
    \end{figure*}
    
    Compared to M20K, the area overhead of M4BRAM-S comes from the duplication shuffler, eFSM, and 4 BPEs, which are 42 $\mu$m\textsuperscript{2}, 238 $\mu$m\textsuperscript{2}, and 856 $\mu$m\textsuperscript{2} respectively. This represents an area increase of 19.6\% over M20K. For \mbox{M4BRAM-L}, the area of BPE and duplication shuffler double \footnote{We do not consider the area overhead associated with memory banking as it is a feature already employed in some commercial BRAMs. The COFFE paper on BRAM \cite{COFFE} reports all results based on 2-bank architecture.}, leading to a 33.4\% area increase compared to M20K. With M20K constituting \textasciitilde29\% of the core area in the two baseline FPGAs, M4BRAM-S and M4BRAM-L will increase the FPGA core area by 5.6\% and 9.5\%, respectively, which are higher than those reported in the previous works \cite{ccb, comefa, bramac} mainly because our baseline FPGAs have a higher M20K:DSP ratio. 
    Furthermore, Section \ref{M4BRAM_vs_DSP} will show that using the same area, M4BRAM will outperform DSP for accelerating mixed-precision DNNs, thus offering a better area vs. performance scaling. 

    The baseline M20K has a maximum frequency of 730 MHz in 20-nm technology \cite{arria10}. With COFFE's 22-nm technology, the BPE's critical path delays are 903 ps and 925 ps in M4BRAM-S and M4BRAM-L, respectively, which are lower compared to BRAMAC (958 ps) because of the smaller dummy BRAM array size. When the BPE is double-pumped, M4BRAM-S and M4BRAM-L will limit M20K's frequency to 553 MHz and 540 MHz, respectively, which are $\sim$1.26$\times$ lower than the baseline. A lower M20K frequency is not a concern because realistic FPGA delays are usually constrained by soft logic and routing, and it is unlikely that an FPGA accelerator will run more than 500 MHz even on Stratix-10 with a more advanced 14-nm technology \cite{hpipeNX}.

    \subsection{Comparison with Prior CIM Architectures} 

    Table \ref{m4bram_vs_prior} shows the architectural differences between M4BRAM and prior CIM architectures for FPGA, including CCB \cite{ccb}, CoMeFa \cite{comefa} and BRAMAC \cite{bramac}. Both CCB and CoMeFa require a transposed BRAM data layout that can not be accessed by DSP, leading to restricted dataflow and limited throughput improvement. 
    Although M4BRAM shares a few common features with BRAMAC, such as using a dummy BRAM array (but with different sizes) to compute MAC2, it can perform mixed-precision computation as opposed to BRAMAC. Furthermore, unlike other works that only support a fixed $N_{I}$, M4BRAM can flexibly adjust its $N_{I}$ based on a DNN's topology to improve the utilization efficiency. \mbox{Notably,} BRAMAC-2SA with $N_{I}$ = 2 can cause significant under-utilization for DNNs that involve matrix-vector multiplication, e.g., unbatched long short-term memory (LSTM) networks or attention-based models that require $N_{I}$ = 1. 
    Finally, M4BRAM only occupies one main BRAM port when performing CIM operations, thus offering more efficient interoperability with DSP compared to other CIM architectures.

    \subsection{Mixed-Precision Accuracy vs. Performance Trade-off}
    \label{Mixed_Precision_Accuracy_Performance}

    \begin{figure*}
        \centering
        \includegraphics[width=1\linewidth]{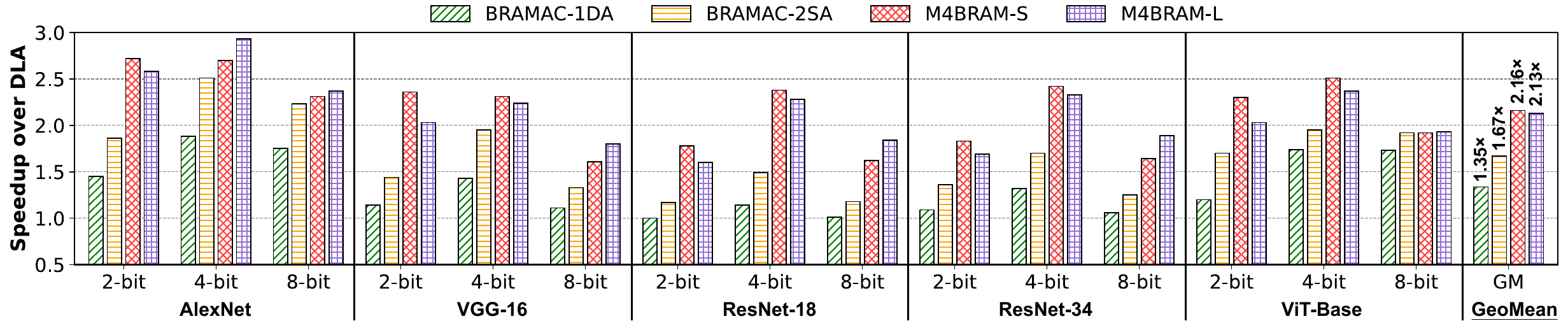}
        \caption{Speedup of M4BRAM and BRAMAC over Intel's DLA for single-precision DNNs.}
        \label{fig10_m4bram_vs_bramac}
    \end{figure*}

    \textbf{Accuracy, performance vs. activation quantization:} Since the computation latency of M4BRAM scales linearly with the activation precision, we first evaluate the model accuracy and performance by sweeping the activation precision from 4 to 8 bits while keeping the weight precision at 8 bits. We choose the GX650 FPGA in this experiment. For Hetero-DLA, we consider three M4BRAM configurations -- M4BRAM-S with double-pumped BPE (DP-M4S), and M4BRAM-L with synchronous (SY-M4L) or double-pumped BPE (DP-M4L). 
    
    Fig. \ref{fig11_m4bram_mixed_activation_performance} presents the Top-1 model accuracy vs. performance trade-off for various DNNs.  
    For each DNN, we observe a drop in speedup when the activation becomes 5 bits, which is due to a 2$\times$ increase in the DSP-packing factor as described in Section \ref{DSP_Packing_Suboptimal}. However, 5-bit activation causes a non-negligible accuracy loss of $>$ 0.5\% for all three evaluated DNNs, with a maximum of 0.94\% for ResNet-18. This highlights that solely relying on DSP-packing cannot provide an optimal accuracy vs. performance trade-off for mixed-precision DNNs. 
    
    When the activation precision is 6-bit, the three M4BRAM configurations deliver an average speedup of 2.16$\times$ across all evaluated DNNs, while incurring an accuracy loss of \mbox{$<$ 0.5\%}. Specifically, the average speedups achieved by DP-M4S, SY-M4L, and DP-M4L are 1.92$\times$, 2.26$\times$, and 2.31$\times$, respectively. When considering cases where the activation precision changes from 8 to 6 bits, the performance of SY-M4L linearly increases, while the double-pumping feature of DP-M4S leads to an incremental speedup for every 2-bit reduction in the activation precision. Interestingly, DP-M4L shows nearly identical speedup. We attribute this to the much higher MAC throughput of DP-M4L compared to DSP, which causes the tile latency to be bottlenecked by the latter.

    \begin{table}
      \centering
      \caption{Top-1 accuracy of ResNet-34 under different \% of 8-bit filters and the speedup of SY-M4L vs. all-4b DLA.}
      \renewcommand{\arraystretch}{1.2}
      
        \begin{tabular}{ m{1.1cm}<{\centering} | m{2.5cm}<{\centering}  m{1.3cm}<{\centering}  m{1.3cm}<{\centering}  m{1.5cm}<{\centering} }
            \toprule
              Model & Weight & Activation & Top-1 & Speedup vs.  \\
              Config & Precision & Precision & Accuracy & 4b DLA  \\
            \midrule
              Baseline & 32b               & 32b & 75.16\%  & N.A. \\
              Retrain  & 32b               & 32b & 75.49\%  & N.A. \\
              Mixed    & 95\% 4b +  5\% 8b & 6b & 75.22\%  & 2.33 $\times$ \\
              Mixed    & 85\% 4b + 15\% 8b & 6b & 75.26\%  & 2.02 $\times$ \\
              Mixed    & 75\% 4b + 25\% 8b & 6b & 75.37\%  & 2.02 $\times$ \\
            \bottomrule
        \end{tabular}
      \label{channelwise_weight}
    \end{table}
    
    \textbf{Accuracy, performance vs. weight quantization:} As described in Section \ref{MixedPrecision_Support}, the MAC throughput of M4BRAM can be scaled by 2$\times$ with a 2$\times$ lower weight precision. This can lead to further performance improvement through intra-layer weight quantization. 
    Recent work has demonstrated that an FPGA-friendly intra-layer quantization approach is to have two filter groups with 4-bit and 8-bit precision, respectively, in every layer \cite{filmQNN}. By properly choosing a uniform activation precision and tuning the ratio \textit{R} of the 8-bit filters, it is possible to achieve higher performance than the pure 8-bit model while maintaining accuracy comparable to the FP32 model. 
    
    To evaluate the potential of Hetero-DLA for intra-layer weight quantization, we choose the GX400 FPGA employing M4BRAM-L with synchronous BPE and follow the approach in \cite{filmQNN} by dividing the FPGA resources into two groups. Each group utilizes M4BRAM-L to store either 4-bit or 8-bit filters and computes convolution using the available DSP and M4BRAM-L resources. We conduct intra-layer weight quantization experiments on ResNet-34 as a case study. Since we have shown that 6-bit activation precision has negligible accuracy loss compared to the FP32 model, we set all activation to 6-bit during mixed-precision training. 
    
    Table \ref{channelwise_weight} shows the Top-1 accuracy of ResNet-34 under different mixtures of 4-bit and 8-bit filters, as well as the resulting speedup achieved by Hetero-DLA over the all 4-bit model on DLA. All mixed-precision configurations achieve $<$ 0.3\% accuracy loss compared to the retrained FP32 model. When the ratio of 8-bit filters \textit{R} increases from 5\% to 15\%, the accuracy increases by only 0.04\%, but the speedup reduces from 2.33$\times$ to 2.02$\times$. This is because with \textit{R} = 15\% 8-bit filters, the number of DSPs on the GX400 FPGA cannot achieve the same tiling configuration as \textit{R} = 5\%. Specifically, the optimal tiling configuration for \textit{R} = 5\% consumes 816 M4BRAM and 612 DSP. To maintain the same tiling configuration when \textit{R} = 15\%, the resource utilization will increase to 912 M4BRAM and 666 DSP, which exceeds the number of DSP available on the GX400 FPGA. These results demonstrate that combining M4BRAM and DSP on a heterogeneous accelerator can provide a good trade-off between accuracy and performance when quantizing both weights and activations 

    \subsection{M4BRAM vs. BRAMAC} \label{M4BRAM_vs_BRAMAC}
    BRAMAC is the state-of-the-art compute-in-BRAM architecture and has outperformed other prior approaches.
    We therefore focus on comparing the performance of M4BRAM and BRAMAC for accelerating real-world DNNs. 
    We follow the same evaluation methodology of BRAMAC \cite{bramac} by using DLA as the baseline accelerator, and calculate the execution time of DLA that employs BRAMAC. 
    We consider precision configurations where weights and activation have the same precision in 2-/4-/8-bit. For VGG-16, ResNet-18, and ResNet-34, the 8-bit configurations are evaluated using the GX650 FPGA due to the need for a larger buffer based on DLA's BRAM usage model \cite{dlaOld}. 
    Other DNNs and precision configurations are evaluated using the GX400 FPGA. 
    In addition, we apply double-pumping to M4BRAM-S to ensure a fair comparison with BRAMAC-1DA since these two architectures have similar M20K area overhead. 
    
    \textbf{Overall performance comparison:} Fig. \ref{fig10_m4bram_vs_bramac} shows the overall speedup over the baseline DLA after replacing the normal BRAM with BRAMAC or M4BRAM. On average, M4BRAM outperforms BRAMAC by 1.43$\times$ across various DNNs. Specifically, BRAMAC-1DA and BRAMAC-2SA achieve an average speedup of 1.35$\times$ and 1.67$\times$ over DLA, respectively. M4BRAM-S and M4BRAM-L provide even higher speedups over DLA, reaching 2.16$\times$ and 2.13$\times$ on average, respectively. These additional performance gains of M4BRAM are attributed to allowing DSP to freely access the main BRAM array and supporting different parallelism configurations. 
    Among the evaluated DNNs, AlexNet and ViT-Base show higher speedup since they contain more output channels in each layer, thus exhibiting higher $N_{W}$ that is well exploited by both BRAMAC and M4BRAM.
    On the other hand, VGG-16, ResNet-18, and ResNet-34 contain relatively fewer output channels in their first several layers.
    As a result, continuing to increase $N_{W}$ as in BRAMAC won't benefit performance due to low hardware utilization, while M4BRAM can exploit weight-sharing (by increasing $N_{I}$) to achieve higher hardware utilization and speedup.
    
    \begin{figure}
        \centering
        \includegraphics[width=1\linewidth]{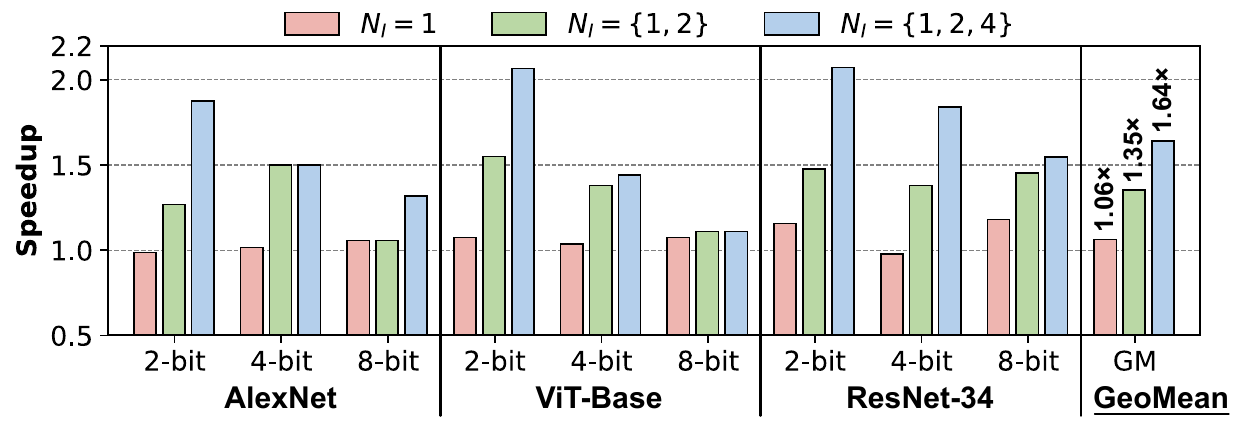}
        \caption{Speedup over BRAMAC-1DA for when M4BRAM-S supports different sets of parallelism configurations.}
        \label{fig11_ablation}
    \end{figure}
    \textbf{Performance ablation study:} The performance benefits of M4BRAM are twofold: better interoperability with DSP by using only one BRAM port for CIM, and flexible parallelism configurations through in-BRAM data duplication. To further investigate the effects of these two features, we conduct ablation studies to compare the performance of double-pumped M4BRAM-S and BRAMAC-1DA.
    We restrict the supported parallelism configurations of M4BRAM by changing the weight-sharing factor $N_{I}$ in our design space exploration tool. To test the performance benefits of M4BRAM's better interoperability with DSP, we simply set $N_{I} = 1$, which effectively disables the in-BRAM data duplication scheme.

    Fig. \ref{fig11_ablation} shows the resulting speedup of M4BRAM-S over BRAMAC-1DA across three DNNs. When $N_{I} = 1$, i.e., without in-BRAM data duplication, M4BRAM-S achieves slightly better performance (1.06$\times$ on average) compared to BRAMAC-1DA. It is important to note that the total dummy array size of BRAMAC-1DA is $7 \times 160$, which provides $1.25\times$ higher peak performance than M4BRAM-S with a total dummy array size of $7 \times 128$. But the better interoperability between DSP and M4BRAM enables a more efficient dataflow, effectively compensating for the lower peak performance. On the other hand, as M4BRAM-S supports more parallelism configurations, the speedup over BRAMAC-1DA becomes higher. When supporting all three parallelism configurations shown in Fig. \ref{fig5_m4bram_parallelism}, M4BRAM-S achieves an average speedup of 1.64$\times$ over BRAMAC-1DA. 
    These results demonstrate the additional performance that is enabled by the flexibility within M4BRAM, despite its lower peak performance compared to BRAMAC.
    
    \subsection{M4BRAM vs. DSP} \label{M4BRAM_vs_DSP}
\begin{figure}
        \centering
        \includegraphics[width=1\linewidth]{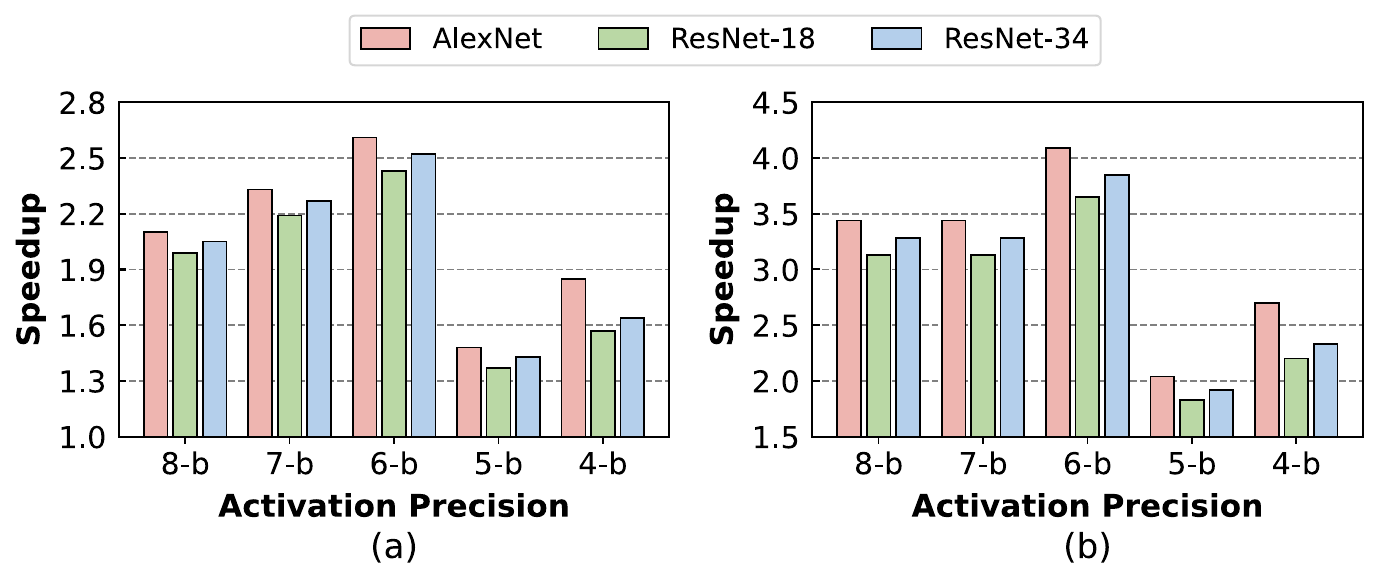}
        \caption{Speedup over GX-DSP when GX-M4 employs M4BRAM-L with (a) synchronous and (b) double-pumped BPE. The weight precision is 8-bit.}
        \label{fig12_m4bram_vs_dsp}
    \end{figure}

    This section shows that M4BRAM can outperform DSP in terms of performance per area. 
    As described in Section \ref{M4BRAM_Area_Frequency}, the area overhead of M4BRAM-L translates to a 9.5\% increase in the FPGA core area. On the GX650 FPGA, this is equivalent to 640 DSPs. To compare the performance of M4BRAM and DSP given the same area budget, we use two GX650-like FPGAs. The first is called GX-M4 with 2489 \mbox{M4BRAM-L} but no DSP, and another is called GX-DSP containing 2489 normal BRAM and 640 DSP. We analytically model the performance of these two FPGAs for accelerating AlexNet, ResNet-18, and ResNet-34 with 8-bit weight. Fig. \ref{fig12_m4bram_vs_dsp} shows the speedup of GX-M4 over GX-DSP when sweeping the activation precision from 4 to 8 bits. On average GX-M4 offers 1.98$\times$ and 2.95$\times$ higher performance than GX-DSP across various DNNs. These results highlight that M4BRAM can offer a better performance vs. area scaling compared to DSP, whose high-precision multiplier suffers from significant under-utilization for low-precision multiplications.

\section{Conclusion} \label{Conclusion}
In this paper, we have proposed a new compute-in-BRAM architecture, called M4BRAM, to enhance the FPGA's capability for accelerating mixed-precision DNNs. M4BRAM supports diverse mixed-precision configurations with 2-/4-/8-bit weights and 2- to 8-bit activations to enable effective performance vs. accuracy trade-off. Evaluation results on various mixed-precision DNNs demonstrate that by employing M4BRAM to a tiled accelerator, an average speedup of 2.16$\times$ can be achieved with negligible accuracy loss of $<$~0.5\% with the mixed-precision DNN compared to the FP32 baseline. Furthermore, M4BRAM offers different parallelism configurations and better interoperability with DSPs, making it outperform a state-of-the-art compute-in-BRAM architecture by 1.43$\times$ on average across various DNNs. Finally, compared to DSPs, we show that M4BRAM can be more efficient for accelerating mixed-precision DNNs. With continuous advancements in mixed-precision quantization of DNNs, we believe that adding M4BRAM to future AI-optimized FPGAs can be highly valuable.

\section{Acknowledgement} \label{Acknowledgement}
We would like to thank the anonymous reviewers for their constructive feedback. This material is based upon work supported by the National Science Foundation under Grant No. 2303626. We would also like to thank Ilya Ganusov for the valuable discussion about FPGA BRAM architectures. 

\clearpage 
\bibliographystyle{IEEEtran}
\bibliography{references}

\end{document}